\documentclass[twocolumn]{aastex61}

\newcommand\ploso{PLoS One}
\newcommand\aastex{AAS\TeX}

\newcommand\natsd{NatSD}
\newcommand\basis{Bulletin of the Association for Information Science and Technology}

\submitjournal{ApJS Special Issue on Data}
\shorttitle{A Model for Data Citation Using DOIs}
\shortauthors{Novacescu et al.}

\begin{document}

\title{A Model for Data Citation in Astronomical Research using\\Digital Object Identifiers (DOIs)}

\correspondingauthor{Jenny Novacescu}
\email{jnovacescu@stsci.edu}

\author[0000-0002-8523-015X]{Jenny Novacescu}
\affil{Space Telescope Science Institute \\
3700 San Martin Drive \\
Baltimore, MD 21218, USA}

\author[0000-0003-4797-7030]{Joshua E.G. Peek}
\affil{Space Telescope Science Institute \\
3700 San Martin Drive \\
Baltimore, MD 21218, USA}

\author{Sarah Weissman}
\affil{Space Telescope Science Institute \\
3700 San Martin Drive \\
Baltimore, MD 21218, USA}

\author[0000-0003-0556-027X]{Scott W. Fleming}
\affil{Space Telescope Science Institute \\
3700 San Martin Drive \\
Baltimore, MD 21218, USA}

\author{Karen Levay}
\affil{Space Telescope Science Institute \\
3700 San Martin Drive \\
Baltimore, MD 21218, USA}

\author{Elizabeth Fraser}
\affil{Space Telescope Science Institute \\
3700 San Martin Drive \\
Baltimore, MD 21218, USA}

\begin{abstract}

Standardizing and incentivizing the use of digital object identifiers (DOIs) to aggregate and identify both data analyzed and data generated by a research project will advance the field of astronomy to match best practices in other research fields like geosciences and medicine. Increase in the use of DOIs will prepare the discipline for changing expectations among funding agencies and publishers, who increasingly expect accurate and thorough data citation to accompany scientific outputs. The use of DOIs ensures a robust, sustainable, and interoperable approach to data citation in which due credit is given to researchers and institutions who produce and maintain the primary data. We describe in this work the advantages of DOIs for data citation and best practices for integrating a DOI service in an astronomical archive. We report on a pilot project carried out in collaboration with AAS Journals. During the course of the 1.5 year pilot, over 75\% of submitting authors opted to use the integrated DOI service to clearly identify data analyzed during their research project when prompted at the time of paper submission.

\end{abstract}

\section{Introduction} \label{sec:intro}

A digital object identifier (DOI) is a persistent, resolvable identifier for an object. When a DOI is resolved through the central DOI service \footnote{\url{https://dx.doi.org/}} it returns current information on that object, including descriptive metadata and an object location \citep{doihandbook2016}. Most researchers in astronomy are familiar with DOIs based on the CrossRef\footnote{\url{https://www.crossref.org/}} schema which are used by publishers to identify individual journal articles. DOIs can also be used to cite data sets as standalone research objects or in relation to one or more publications.

Three examples of services which allow researchers to build and share data sets are \href{https://zenodo.org/search?page=1&size=20&type=dataset}{Zenodo}, 
\href{https://dataverse.harvard.edu/}{Dataverse}, and \href{https://figshare.com/}{figshare}. Zenodo is part of the OpenAIRE project which was ``commissioned by the EC [European Commission] to support their nascent Open Data policy by providing a catch-all repository for EC funded research.\footnote{\url{http://about.zenodo.org/}}''. A number of major journal publishers are already integrated with figshare to host large amounts of data linked to online articles. Zenodo, Dataverse, and figshare all provide the ability to assign DOIs to deposited data and research outputs. Through use of DOIs, data creators and authors can ensure primary data analyzed as well as  data derived and generated in the research process are easy to find, attributable, and accessible well into the future.

Once a DOI is assigned to a data set it can be included in the manuscript of a journal publication, associated with the publication's (CrossRef) DOI, and its metadata made available in public databases for ingestion into discovery systems like ADS. 

We are proposing that astronomical archives, as the hosts and curators of data, consider integrating a similar DOI workflow. Greater adoption of DOIs for archival data, especially in reference to complex data sets, creates a stable data environment in which the reader is taken to the primary observations or derived data referenced in the literature, no matter its current location online or the length of time since publication. Accessing data and building data sets via an archive-integrated DOI service allows the researcher to locate, combine and then subsequently cite data whose provenance and relation to other observations would not be as apparent if the data set existed as a discrete research object external to the archive.

Typically, services like Zenodo, Dataverse and figshare provide data hosting for researchers to share data sets that would otherwise not be available to the general research community. In the case of astronomical research, it is more common to combine or refine data that is already hosted and shareable via a public data archive. Four examples of public data archives are the Barbara A. Mikulski Archive for Space Telescopes (MAST)\footnote{\url{https://mast.stsci.edu/portal/Mashup/Clients/Mast/Portal.html}} at the Space Telescope Science Institute (STScI), the Centre de Donn\'ees astronomiques de Strasbourg (CDS)\footnote{\url{http://cds.unistra.fr/}}, the Canadian Astronomy Data Centre (CADC)\footnote{\url{http://www.cadc-ccda.hia-iha.nrc-cnrc.gc.ca/en/}}, and the High-Energy Astrophysics Science Archive Research Center (HEASARC)\footnote{\url{https://heasarc.gsfc.nasa.gov/}} hosted at the NASA Goddard Space Flight Center.

The challenge is how best to provide accurate identification of the data analyzed without needing to duplicate large amounts of data or reprint large tables of observational metadata already available in these public databases. The additional challenge is how to share data that was further processed or refined by the research team (hereafter referred to as ``data generated''), but which originated in the archive or is otherwise connected to the missions hosted by the archive.

The Space Telescope Science Institute (STScI), which is operated by AURA for NASA, was founded in 1982 to oversee the scientific and data archiving operations of the \emph{Hubble Space Telescope} (HST). STScI will also manage the flight and science operations of the \emph{James Webb Space Telescope} (JWST). MAST, hosted at STScI, collects and provides access to astronomical data from over 20 missions, with a historical focus on data in the optical, ultraviolet, and near-infrared parts of the electromagnetic spectrum. Archived data comes from missions including {Hubble}, Kepler/K2, GALEX, and FUSE. Future missions like TESS and WFIRST will also house their data in MAST.

In consultation with the American Astronomical Society journals, STScI has been working on a pilot project to provide authors a means to assign a DOI to the MAST data set referenced in their original research articles. AAS was chosen as the initial partner for this pilot project because a high percentage of papers which cite MAST and specifically HST data are published in AAS journals, including Astronomical Journal (AJ), Astrophysical Journal (ApJ), and Astrophysical Journal Supplement (ApJS). Of the over 15,000 papers identified since 1991 that use HST observational or archival data from MAST, over 55\% were published in AJ or ApJ. Archival data accounts for over half of all publications related to the Hubble Space Telescope.

The pilot project was initiated in late 2015 and launched in early 2016. It allows authors to locate DOIs for predefined data sets (called High-Level Science Products\footnote{\url{http://archive.stsci.edu/hlsp/index.html}}) or to create their own data sets using the MAST DOI Portal tool\footnote{\url{https://mast.stsci.edu/portal/Mashup/Clients/DOI/DOIPortal.html}} developed by Weissman and Tom Donaldson (STScI).

During the current pilot phase, the ability to create a custom DOI is available only to authors affiliated with STScI at the time of submission. Plans are underway to expand the service to 18 additional institutions which have historically produced a high percentage of publications that use MAST data.

In \S \ref{sec:doidev}, we discuss the development of DOIs as a concept and interoperable object identifier. \S \ref{sec:datacitation} addresses data citation principles in the scholarly community, while \S \ref{sec:datacitesci} provides specific examples of DOI use in other scientific fields. In \S \ref{sec:mastaasprinciples}, we discuss the data citation principles and DOI implementation goals mutually agreed upon by STScI and our AAS journal collaborators at the start of the pilot project. \S \ref{sec:doicreation} and \S \ref{sec:mastdoiresolution} outline the user experience, both in terms of creating a DOI and resolving a data DOI referenced in the literature. Pilot project outcomes are reported in \S \ref{sec:pilotoutcomes}. Finally, we close with \S \ref{sec:techchallenges} and \S \ref{sec:pilotandgoals} to outline some of the lessons learned and future directions of the pilot project.

\section{History of DOI Development} \label{sec:doidev}

The use of DOIs to capture sets of data has been proposed in existing principles, manifestos, standards documents, journal articles, and best practices across scientific fields since at least the early 2000s \citep{edmunds2012,paskin2005,callaghan2012}. 

Before discussing current applications of DOIs for data citation in research areas outside astronomy, it is important to understand the development of the DOI as a concept. The structure, assignment, creation/registration of DOI names, resolution, and interoperability standards for DOIs are codified by the International Standards Organization in ISO 26324:2012\footnote{\url{https://www.iso.org/standard/43506.html}}. The International DOI Foundation, which governs the DOI system, lays out the three most essential properties of DOIs\footnote{\url{https://www.doi.org/factsheets/DOIKeyFacts.html}}. DOIs are:

\begin{itemize}
\item Actionable---through use of identifier syntax and network resolution mechanism (Handle System\textregistered)
\item Persistent---through combination of supporting improved handle infrastructure (registry database, proxy support, etc) and social infrastructure (obligations by Registration Agencies)
\item Interoperable---through use of a data model providing semantic interoperability and grouping mechanisms 
\end{itemize}

There is no means through which one can update or even identify all research publications or data products that point to a specific URL. As the location of a resource changes via edits to the URL, it can become difficult or impossible to find cited data without proper redirection. \citet{paskin2005} explains in detail how DOIs are superior to URLs in this regard. 

Earlier research \citep{2014PLoSO...9j4798P} has demonstrated that over 40\% of data links provided via traditional URLs in the astronomical literature are broken within a decade of publication, and over 10\% of traditional links become invalid within just three years. By registering a DOI it is possible to minimize this problem, as the publication will reference a digital identifier which carries out the work of locating the object online. DOIs are editable and can be updated via an API provided by the registration agency when the associated URL(s) changes. This coupled with the central resolution service provided by dx.doi.org ensures that DOIs are actionable and persistent. In the most common scenario, a DOI is created to resolve the location of a resource, typically via the URL assigned to that digital object, whether a research data set or journal article.

Institutions wishing to create DOIs for persistent, actionable links to data must first enter into an agreement with a registration agency (RA) such as DataCite, the most well-known global registration agency for data DOIs. \citet{edmunds2012} provides a thorough history of DataCite's role in the modern research landscape. A valid registration agency, as regulated by the International DOI Foundation (IDF), has both the necessary infrastructure and rights to create and register a DOI. RAs may implement the metadata scheme of their choice, but the IDF specifies a minimum metadata standard---the DOI metadata kernel---which ensures a level of interoperability between DOIs issued by different agencies. Furthermore, DataCite has been capturing mandatory metadata associated with \emph{all} DOIs since 2011, including those created by other IDF-accredited registration agencies. 

By defining mandatory and recommended properties and values in a standardized metadata schema, it is possible to convey a number of important elements about the data and its provenance, including the data producers' names and affiliations. The publisher or holder of the data (in our case MAST) and resource type are also identifiable, as is an optional contributor value which allows acknowledgement of those who in some way contributed to the data (perhaps by designing a software package to process the data in novel ways) but did not play a part in the data's original creation. Other elements such as version, funder, and associated project number are available for use according to the widely-used DataCite Metadata Schema\footnote{\url{https://schema.datacite.org/}}.
 
In MAST's case, many of these optional fields are not used at this time because a) we are in the process of migrating from one registration agency to another and are working towards DataCite indexing compliance for existing DOIs, and b) essential metadata values are already associated with observations within the archive and can be easily found when accessing the data set via a resolved DOI. It should be noted, however, that any DOI value or property an archive wishes to make searchable must be encoded in the DOI metadata in order for those properties to be exposed in discovery systems such as \href{https://search.datacite.org/}{DataCite.org}, ADS, or Google.

The goal of the pilot was to provide a means to reference large and complex data sets containing tens or thousands of observations. In one instance, a submitting author created a DOI to reference more than 14,000 rows of MAST data. Without a DOI to succinctly group a large volume of observations, such data representation would be limited in a traditional journal article and the rich metadata associated with the data set limited if deposited elsewhere online, separate from the archive. 

\section{Data Citation Principles} \label{sec:datacitation}

A number of guidelines and standards have already been developed which support the concept of DOIs as a means to reliably and accurately cite data. The FAIR Data Guiding Principles \citep{2016NatSD...360018W} advise that data must be \textbf{F}indable, \textbf{A}ccessible, \textbf{I}nteroperable, and \textbf{R}eusable in order to be of enduring value to the science community. The FAIR Data Principles were developed by FORCE11, the Future of Research Communications and e-Scholarship\footnote{\url{https://www.force11.org/}}, an open organization that seeks to provide guidelines for best practices in scholarly communication in the digital age that benefit researchers, institutions, data archives and repositories, journal publishers, libraries, indexers, and other stakeholders in scholarly communications.

Because astronomers have a common experimental platform in the shared sky, they can observe the exact same coordinates while leading unrelated investigations, and often benefit from complimentary archival data obtained from observations of the same target or field. For these reasons, there is already a large emphasis on interoperability in the field of astronomy. This is apparent from the twice yearly virtual observatory (VO) interoperability meetings and the numerous VO standards\footnote{\url{http://www.ivoa.net/}} adopted by astronomy data centers worldwide. Due to the inherent cost in obtaining observational data and limits to available observing time, data preservation and accessibility are also inherent concerns in astronomy.

In the context of MAST, data is FAIR at the individual observation level via a well-maintained public interface using standard formats and data models like FITS and the Common Archive Observation Model (CAOM). The primary challenge is how to meet FAIR standards for potentially large and seemingly arbitrary collections of observations, in particular when a future researcher wishes to refer to or reuse the data.

It is the reusability principle and its connection to data provenance that most concerned the DOI pilot project group. Conscientious use of DOIs and an associated metadata schema, as discussed in \S \ref{sec:doidev}, can align data citation practices in astronomy to many of the guiding principles presented by \citet{groth2013} in the W3C PROV documents\footnote{\url{https://www.w3.org/TR/prov-overview/}}. The PROV Family of Documents consists of a set of recommendations for describing provenance between raw data and other entities such as combined data sets (DOIs), new data generated through computational processing of observational data, or scientific publications that exist in an open, heterogeneous environment like the Web. 

Discussions surrounding data citation standards and practices in astronomy are not as well-developed as they are in other disciplines such as genomics and environmental sciences \citep{edmunds2012,callaghan2012}. Expectations for accurate and concise attribution of data remain inconsistent across institutions and publishers in astronomy, which is turn means data sets are still difficult to find even if published \citep{2015BASIS..41...40H}.

In STScI's case, linking publications to precise data sets and observations can be a challenge. At times, it is not possible to determine which archival data were analyzed because of the lack of identifying information or errors in data citation in the publications. For the HST bibliography, slightly more than 5\% (n=763 of 14,459 publications through 2016) cannot even be attributed to a specific grant. The other 95\% required significant human effort (1 FTE) to identify MAST observations, and in many cases only the associated grant is known, not the individual observations.

\section{Data Citation in Other Sciences} \label{sec:datacitesci}

To solidify the case that data citation matters to scientific advancement and is becoming an expectation across the spectrum of research fields, we briefly summarize the role of data citation in other sciences, such as biomedical research, climatology and earth sciences, and oceanography.

In the field of medicine, data citation is required in some cases, although not yet incentivized across all medical fields. As part of 21st Century Cures Act \footnote{\url{https://www.congress.gov/bill/114th-congress/house-bill/6}} passed in December 2016 by the U.S. Congress, some types of National Institutes of Health (NIH) funding, especially that linked to clinical research, are contingent on data disclosure and sound data management and citation. From \citet{bierer2017}:

``Data sharing, whether elective or required, creates an obligation for the original investigators who obtain funding, design studies, collect and analyze data, and publish results to make their curated data and associated metadata available to third parties''...``Data from well-designed and well-executed research not only are useful for the original purpose and secondary analyses by the original researchers but also can be repurposed for a variety of applications, including independent replication, avoidance of duplicative studies, generation or testing of new hypotheses, and the general advancement of clinical and biologic understanding.''

Although Bierer is referencing biomedical data, the basic premise that accurate and thorough data citation has implications for future research can be extended from medicine to any other scientific field, including astronomy. \citet{force11} expresses a similar emphasis on data citation for research integrity and permanence in their Joint Declaration of Data Citation Principles.

Examples of data archiving and data citation standardization in other sciences are illustrated by the NOAA National Centers for Environmental Information (NCEI)\footnote{\url{https://www.ncdc.noaa.gov/about/}}, the USGS ScienceBase Catalog\footnote{\url{https://www.sciencebase.gov/catalog/}}, and the NASA Climate Model Data Services\footnote{\url{https://cds.nccs.nasa.gov/}}, all of which employ DOIs to resolve hosted data sets. The USGS in particular has some of the most robust examples of enhanced DOI resolution and metadata-enriched DOI landing pages\footnote{\url{https://doi.org/10.5066/F7V98691}} \footnote{\url{https://doi.org/10.5066/F7FT8J5T}}, and could serve as a model to astronomy.

Many of these data repositories arose in their respective fields because there was no shared data archive, or solely because government mandates require federal agencies to make their data available to the public. In this sense observational astronomy is unique because the international community has voluntarily taken measures to capture and standardize data formats across a broad range of databases including the NASA/IPAC Extragalactic Database (NED), SIMBAD, and various virtual observatory (VO) compliant archives such as MAST, IRSA, and HEASARC.

The case for more stringent data citation in astronomy is strengthened when considering the increase in the quantity and complexity of data from survey telescopes, beginning with the Sloan Digital Sky Survey (SDSS) \citep{york2000}, continuing with ESO's VLT Survey Telescope (VST) and others, and moving forward with the Large Synoptic Survey Telescope (LSST) and the Wide-Field Infrared Survey Telescope (WFIRST)\citep{2014PLoSO...9j4798P,zhang2015,2015BASIS..41...40H}. 

Moreover, standardizing data citation practices across institutions and archives will become increasingly critical so researchers can more easily cite and provide access points to data from multiple archival repositories in a single study. A basic ADS search for abstracts noting ``multi-mission'',``multi-wavelength'', or ``multi-messenger'' reveals a sharp increase in the use of combined observations since 2000. ADS references 1.7K publications which mention multi-mission or multi-wavelength studies in the abstract or title within the astronomy and astrophysics collection during the 28 year period of 1972-1999. For the 17 year period from 2000-2016, there are over 11.6K ADS records matching the same search criteria at the time of this writing.

\section{MAST---AAS Joint Principles on Digital Object Identifiers} \label{sec:mastaasprinciples}

Considering the shared goals of accurate and unambiguous data citation, MAST and AAS Journals developed the following mutual principles when the DOI pilot project began in late 2015:

\begin{enumerate}
\item \textbf{Journal publishers and archives must actively solicit DOIs for data identification at the time of manuscript submission}.

The archive must provide a simple and elegant tool within its interface if authors are expected to use the service. Consolidating multiple observations into a single DOI created via the archive has the advantage of limiting formatting issues for both author and publisher and satisfies the author's need to host potentially large data sets for future reference and the reader's need to resolve to a persistent location years later.

STScI and AAS agreed that in the ideal standard, a researcher's supporting data (both primary data analyzed and data generated or derived as part of the research) should be submitted at the time of publication so the DOI can be persistently linked within the manuscript and, ideally, its associated metadata upon final publication. This matches the recommendations made in the FORCE11 Data Citation Roadmap for Scientific Publishers \citep{cousijn2017} \emph{[in progress]}.

In cases where data set identification is not completed during submission but the author later indicates a desire to publish a data set DOI in relation to the manuscript, it is possible for STScI to create a DOI and communicate the relevant metadata to AAS for inclusion if final manuscript formatting has not occurred. 

\item \textbf{Minimal integration between archive and publisher is best}. This model allows for interoperability among other publishers/archives in the future.

DOIs are communicated from MAST to AAS by simple cut-and-paste into a web form. There is no hand off of data or metadata happening behind the scenes. This has the major benefit of easy federation and few additional standards. All metadata associated with the data DOI is retained and controlled by the archive. All metadata associated with the publication is controlled by the publisher.

The expectation is that AAS Journals will work with other archives to integrate DOIs into articles and MAST will partner with other journals in the future. When initiating the pilot, we could see no motivation for a more complex system.

\item \textbf{Fixed DOIs should be made available}.

Fixed DOIs refer to High-Level Science Products (HLSPs) and defined subsets of mission data such as entire quarters of Kepler data. In many cases, these large data subsets are referenced as a whole in a manuscript and used throughout the course of research. Authors have the ability to identify HLSPs or defined data subsets using pre-assigned, fixed DOIs.

High-Level Science Products (HLSP) are observations, catalogs, or models that complement, or are derived from, MAST-supported missions. HLSPs can include images, spectra, light curves, maps, source catalogs, or simulations. They can include observations from other telescopes, or MAST data that have been processed in a way that differs from the primary data available in the archive. HLSPs are permanently archived at MAST, get their own project webpage, and appear in MAST search interfaces along with bibliographic references to the publications which cite them. 

Having a menu of fixed DOIs available for author selection eliminates the problem of redundant DOIs to refer back to the same data product or subset of mission observations. By assigning consistent identifiers to High-Level Science Products and frequently used data subsets, MAST has provided authors with a persistent link to the data and encourages accurate citation and proper acknowledgment, thus building a clearer picture of the data's reach and contribution to the field.

\item \textbf{Custom DOIs should be made available}.

In contrast, an author may need to refer to sets of previously unrelated observations. Custom DOIs allow the author to compile and identify observations which might otherwise appear unrelated into a single data set using the MAST DOI Portal tool. 

Authors are given flexibility to create one DOI for all observations analyzed in a research publication or multiple DOIs for different sets of observations (grouped by filter, target, or wavelength, for example). The decision to represent data sets as a whole or as subsets depends on how the author feels the research process is best represented.

\item \textbf{Data DOIs refer to first-class research objects, but are not first-class \textit{citable references} on their own}.

This principle is the most difficult to negotiate and is the one that may generate the most conflict in the astronomy and general scientific community. While DOIs are often used to link to and identify first-class research objects, e.g. research articles or data sets, MAST and AAS are in agreement that data DOIs should not be treated as first-class \textit{citable references} on their own, and thus should not show up independently in the bibliography. Our concern is that even if the relationship between the data set and earlier publications which analyzed, refined, or generated the data are retained through proper use of PROV metadata, separating the data from an author's interpretation in an earlier publication can impact how the data is assessed or reused in future studies. As \citet{lawrence2011} explains, the data consumer will often need the accompanying research article and summary of earlier conclusions to make sense of the data. In MAST's context, data DOIs are considered first and foremost digital identifiers for data and metadata packages with the potential to show a relationship between the data and a publication. 

This approach varies from the FORCE11 Joint Declaration of Data Citation Principles and the FORCE11 Data Citation Roadmap for Scientific Publishers \citep{cousijn2017} \emph{[in progress]} but does not reject the basic idea that raw data and data sets are research objects. Part of the argument for more stringent data citation is to credit the creators of data \citep{bierer2017,{cousijn2017}} and MAST agrees this is one of the goals behind the DOI pilot project. In order to do this, however, the astronomy community must come to some agreement on which parties should be overtly credited as creators or contributors of a data-oriented research object.

MAST DOIs referencing sets of analyzed data usually have ambiguous ``creators" and ``contributors". The PIs and Co-Is of the original projects, software engineers who develop a novel program or script to process data, and the authors who carefully curate a heterogenous set of data can all lay claim as creators or contributors of the data set. Thus, ``authorship" for these data sets in the sense that astronomers use it for all other cited references in bibliographies is not yet well defined. In our current model, only the submitting author of the manuscript who combines MAST observations into a set of analyzed data during the submission process is identified with the \emph{Creator} property in the metadata schema describing the DOI. 

For data generated during the research process and later identified with a DOI, authorship is more clearly defined and would typically match those who author the manuscript. In this case, we would again purport that the manuscript itself is the appropriate entity to cite in the bibliography until the astronomical community reaches a consensus on how best to represent data authorship. Encouraging discussion on how creators and contributors of astronomical data can be recognized via standardized, provenance-based metadata, as well as which parties should be permitted to lay claim as a creator or contributor, is one of our motivations for sharing our current model with the community.  

\item \textbf{DOIs refer to the described data set}.

Because the intention of the DOI is to describe the data set analyzed in the paper, we do not force the DOI to be forever fixed in the case where the author made some kind of mistake early on in DOI generation. Our protocol is for MAST to check with the AAS before changing the content of the data set to which the DOI is linked, but we generally believe that it makes sense to allow users to edit their DOIs via a mediated process to match the content of their papers while the paper is still in the typesetting process.

Although we have not yet encountered the situation during the pilot, the expectation is that data DOIs associated with a published manuscript would be subject to version control using the options \textit{isNewVersionOf} and \textit{isPreviousVersionof} in the \textit{relationType} property. This would indicate that a DOI is supplanted by a corrected or updated version.

Similar discussions surrounding how to handle versions of fixed DOIs are ongoing at MAST. Newer versions of catalogs and other products or data subsets are released periodically. In an effort to comply with the DataCite Schema it may be preferable to issue a new DOI per version while still making all earlier versions accessible as we move forward with the DOI pilot.

Per International DOI Foundation standards, there is no way to literally delete a DOI once it has been created. A identifier created in error or found to be spurious can be set to direct the user to a new DOI created in its place, with the appropriate relationship to the original DOI shown in the metadata. Full control over DOIs issued by an archive remains with the archive. This ensures both persistence inherent in DOIs and research integrity.

\end{enumerate}

\section{Process for Creating a DOI During Paper Submission to AAS} \label{sec:doicreation}
In an effort to promote data citation via DOIs for other archives, publishers, and stakeholders we are outlining the simple workflow we have established in coordination with AAS Journals.

\begin{figure*}
\centering
\includegraphics[width=9cm]{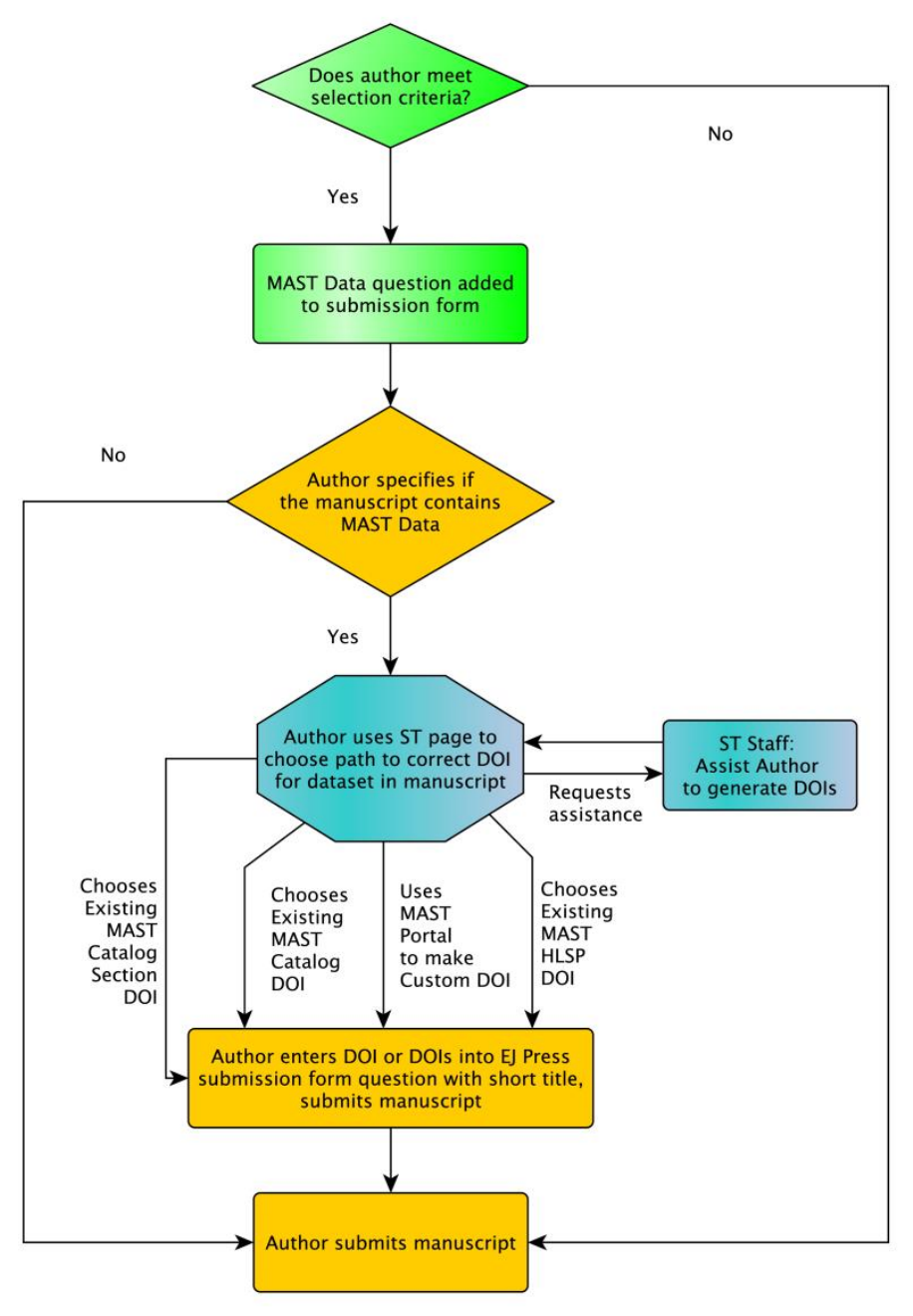}
\caption{Diagram outlining author interaction with the eJournal Press AAS Journals submission page and MAST. Green indicates eJournal Press site functions; yellow indicates author actions on eJournal Press site; blue indicates actions on MAST site.}
\label{fig-1}
\end{figure*}

\begin{enumerate}
\item The author begins paper submission on the E-Journal (EJ) Press website to submit to AAS titles. The EJ Press submission form asks whether data from the MAST archive was used in the manuscript. 

At this time, only submitting authors whose email domain ends in @stsci.edu are prompted. STScI is currently working with the AAS to expand the domains and invited institutions. In an effort to roll out the DOI service in a controlled manner and to understand any technical or conceptual pitfalls before opening the DOI service to a wider audience, we wanted to trial the service with local staff. Advantages of a controlled release are noted in \S \ref{sec:techchallenges}.

\item The author specifies whether MAST data was used. ``No'' reroutes the author back to finish the paper submission process on EJ Press. ``Yes'' routes the author to a MAST DOI home page\footnote{\url{http://archive.stsci.edu/doi/search/}}. 

The intention is to have the author use the MAST-provided DOI service to identify MAST data, rather than posting data tables on personal research pages or creating a DOI through a non-affiliated service. We have yet to experience a situation where the researcher has already used another DOI service such as Dataverse or figshare to compile and cite their data set prior to manuscript submission, though we need to keep in mind this could impact a researcher's willingness to use the MAST service if they feel it is repetitive or don't understand the value in creating a DOI via the archive itself. 

\item From the MAST DOI home page, authors are asked if they used:

a) a collection of specific observations (custom DOI);

b) data from a High-Level Science Product (fixed DOI type);

c) a catalog, e.g., Kepler/GALEX (fixed DOI type); or

d) a large, but well-defined subset of mission data, e.g. a quarter of Kepler long cadence data (fixed DOI type).

\item Authors who select option \textit{a} are directed to the custom version of the MAST DOI Portal to aggregate individual observations used in their research. When creating a custom DOI, the author submits basic metadata such as their name, data set title, and free-text data set description.

Other metadata, such as date created and data set identifier are auto-assigned. As discussed in \S \ref{sec:doidev}, STScI is investigating expansion of DOI metadata fields, though most relevant metadata about the individual observations such as unique identifier, instrument, and wavelength are already stored in MAST and are not replicated within the DOI metadata. 

Authors who select options \textit{b}, \textit{c}, or \textit{d} are prompted to select a pre-assigned, fixed DOI(s) from a list.  As noted, authors have the liberty to mint a single DOI for all observations or a subset of DOIs for different sets of data. Alternatively, if an author referenced an HLSP or other pre-defined subset in their research (fixed DOI) and also analyzed individual observations from the MAST Portal, the individual can submit multiple types of DOIs with his/her manuscript.

The author receives an automated email with a summary of the DOI metadata for each DOI created.

\item Once a custom DOI(s) is created and/or a fixed DOI is selected, the author is taken back to the EJ Press submission form where they cut and paste the DOI(s) and complete the paper submission.

At this point, the author has completed the process.

\end{enumerate}

\section{Resolving DOIs for MAST data sets} \label{sec:mastdoiresolution}

To better understand the user experience from a reader's point of view, we are providing screenshots of  DOI resolution in action. 

When a user links from a data DOI referenced in an article or from a federated search interface such as DataCite.org or ADS, they are directed to a landing page where basic metadata about the data set or data product is provided. In the first example, the user is led to a custom data set within the MAST Portal where they can inspect, download, and otherwise interact with the custom data set. In the second example, the user is led to the landing page for a fixed DOI where further information is given about the data product and where they can view associated data. \textbf{See figure 2 and figure 3}.

\begingroup
\begin{figure*}
\centering
\includegraphics[width=12cm]{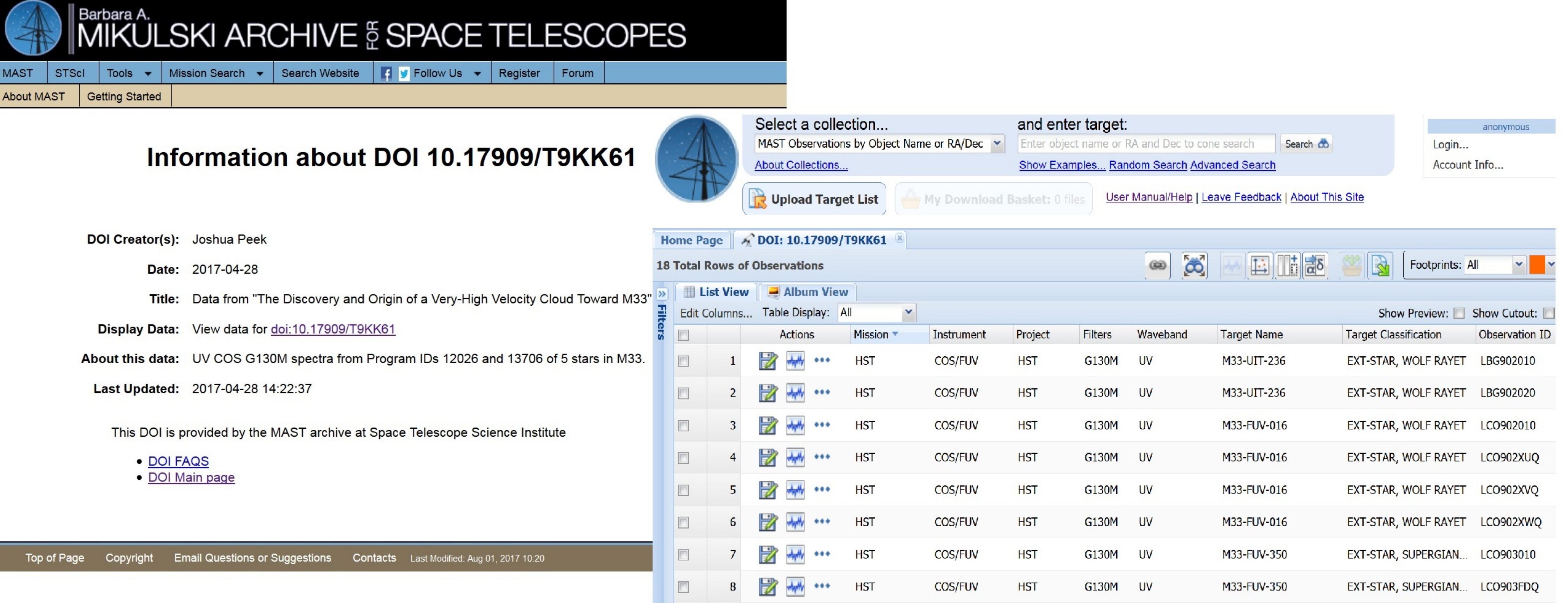}
\caption{Image of DOI resolution page (landing page) for a custom DOI, \url{https://dx.doi.org/10.17909/t9kk61}. Includes basic metadata and link to specific observations within the MAST Discovery Portal.}
\label{fig-2}
\end{figure*}

\begin{figure*}
\centering
\includegraphics[width=12cm]{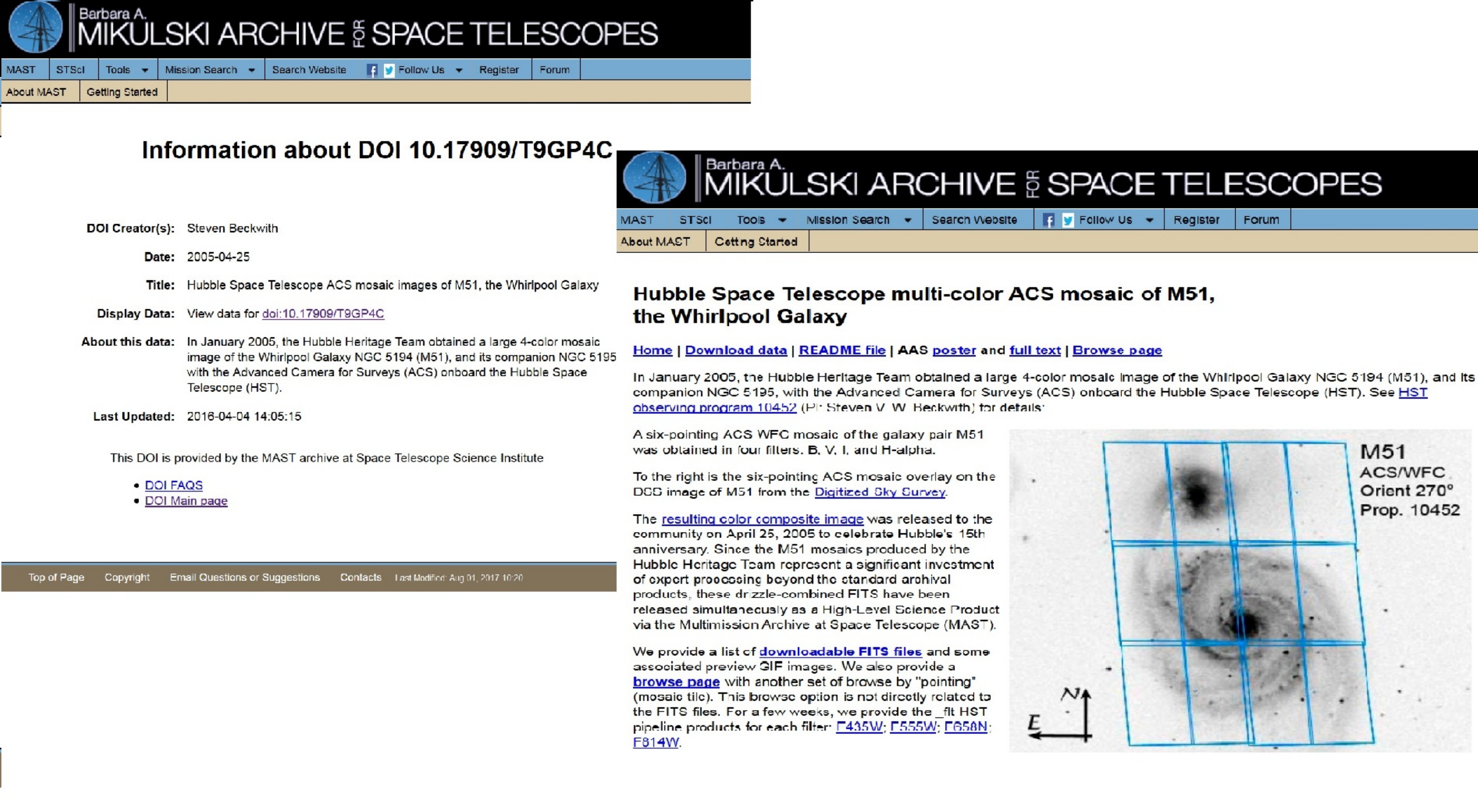} 
\caption {Image of DOI resolution for a High-Level Science Product DOI, \url{https://dx.doi.org/10.17909/t9gp4c}. Landing page includes basic metadata. Link takes user to additional information on data provenance and images.}
\label{fig-3}
\end{figure*}
\endgroup

\section{Pilot Project Outcomes} \label{sec:pilotoutcomes}

A metric demonstrating whether authors were in favor of data citation via DOI and understood the process was possible only because AAS Journals staff committed to tracking papers submitted by corresponding authors with an \textit{@stsci.edu} domain and verifying whether the manuscript analyzed data originating in MAST. Because, ethically speaking, the pilot project group at STScI was not able to view or assess submitted manuscripts, it was necessary to rely solely on the journal publisher for this metric. This level of publisher investment may not scale well if we wish to expand to other journal publishers.

We were able to determine that there has been a 77.2\% compliance rate for DOI creation. During the 1.5 year course of the pilot project, 17 DOIs were created or selected out of 22 eligible submissions. Fifteen of the 17 were custom DOIs for which the author aggregated observations from the MAST DOI Portal, and the remaining two of the 17 submissions selected fixed DOIs to attribute High-Level Science Products. One submission created two custom DOIs for associated data sets but was only counted once in order not to skew the percent in compliance on this small-scale pilot project.

\section{Implementation Challenges and Technical Considerations} \label{sec:techchallenges}

The primary challenge in implementing a DOI service within an astronomical archive is user education. Discussions on the importance of stable resolution to \emph{FAIR} data, the usefulness of unambiguous data set identification to the reader, and its value to the submitting author, data creator, and archive itself are not easily conveyed via a web form in the middle of the paper submission process.

Coming to a consensus within the astronomical community on how we define data attribution and assign value to data creators in the same way we define metrics for citations to refereed literature is a challenge that must be addressed while DOI integration for data attribution is still a relatively new concept. As noted, other sciences such as biology and medicine \citep{edmunds2012,force11,bierer2017} are beginning to place as much emphasis on data creators and seamless data references as the derived research findings reported in traditional refereed articles.

In the FORCE11 model, data associated with a researcher would be easy to find--- both data sets analyzed in their publications and also data generated or derived from others' observations---with a distinct attribution to those responsible for the primary data.

Regarding other implementation challenges in our current model, submitting authors are expected to self-report when asked whether their manuscript used MAST data. This alone can create confusion and already MAST intends to revise its original question on the EJPress website from:\\
\textit{``Does your manuscript directly refer to data in MAST (i.e., data from Hubble, Kepler, GALEX, IUE, etc.)?''}\\
to:
\\
\textit{``"Does your manuscript use or analyze data from Hubble, Kepler, GALEX, IUE, or other data in MAST?''}

Authors who encountered the original question recommended a change to the wording. 

By choosing to implement a DOI creation service in MAST for STScI authors first, the DOI pilot project members were able to follow up with those who did not self-report to determine if the authors did not create a DOI because they misunderstood the question on the EJ Press platform, or because of some other issue with the DOI model. Reasons for non-compliance included confusion over the purpose of DOIs, concerns with how data would be hosted and permanently maintained, confusion over the provenance of data, or a perceived time investment during submission. For this reason the pilot project group decided a third option for \textit{``Yes, but I need assistance''} may be warranted in the expanded pilot. Submitting authors who select this option would then fill in a brief ``contact us'' form directed to MAST and the STScI Library staff.

Even among a pilot user group limited to STScI submitting authors, there was some confusion as to which types of data originated in MAST and qualified for DOI-creation. The other consideration is MAST data from missions hosted in other repositories. If the author obtained the data originating in MAST from a different repository, is it reasonable to ask the author to create a DOI using the MAST DOI service or would DOI creation mediated by the other repository be more logical so long as the MAST-provenance was clear? Adding another option besides a straightforward yes/no and encouraging the author to contact us in such situations will help until there are robust, shared data citation policies and best practices across astronomical archives.

With regards to technical challenges, a major concern is how to standardize metadata schema properties so DOI creation is scalable and automated to the extent possible for astronomical archives, and so DOIs can be properly mined by indexing services such as DataCite and ADS. Keeping in mind that a simplified user experience and data discovery are also end goals, a data set is only as easy to create and find as the metadata properties allow it to be, so negotiating which properties will be considered mandatory, recommended, or optional among astronomical archives is a process that must begin as more archives in addition to MAST consider adopting the DOI model.

Refining the MAST DOI Portal tool and integrating it more fully with the rest of the MAST suite of data search tools is another technical challenge that MAST developers continue to face. Because scientists often alter data in a significant way from its original downloaded form, renaming or combining files, or processing data via customized software and simulations, it can be challenging to recreate a set of observations as it originally existed in the archive portal. Tools to help researchers more quickly identify the primary data they used are in future development plans. Examples include allowing users to save the contents of their carts and to view download histories.

MAST also recognizes that archive users must be able to create and save sample data sets that capture only specific spectra, wavelengths, subclasses, etc. within a larger catalog, atlas, or other science product. Otherwise, the issue of unambiguous data citation is only partially resolved. Further integration of the data that makes up High-Level Science Products and catalogs within the MAST Portal will partially eliminate this problem, but there may be some types of derived data or science products, such as models, that cannot be integrated within the traditional archive infrastructure. 

Being able to show that a subset is part of a larger data set via the relationship between DOIs in the metadata schema, and being able to delineate older and newer versions are both technical issues that are yet to be resolved. 

From the publisher's standpoint, once a DOI is communicated in the manuscript, the publisher must find a way to express the data DOI in the markup version. \aastex6.1 uses their existing \textit{dataset} tag to identify data DOIs, though at this time data DOIs are not embedded in the metadata for the publication itself. This means the relationship between the data analyzed (DataCite DOI) and the article (CrossRef DOI) are not clearly expressed. Following the advice of \textit{A Data Citation Roadmap for Scientific Publishers} \citep{cousijn2017} \emph{[in progress]} and defining this relationship more clearly will be essential if the astronomical community intends to align itself with data citation best practices. 

From the indexing standpoint, data curators and publishers must provide sufficient metadata to make the data discoverable within indexes like ADS and DataCite and coordinate with these indexers if all stakeholders in astronomy and astrophysics are to reap the full benefits of data citation via DOIs. Until clear data attribution is mandated and enforced by granting institutions and funders, it will require a continued conversation on the purpose and goals of data attribution and its value to the astronomy community to encourage greater compliance.

\section{Pilot Expansion and Long Term Goals} \label{sec:pilotandgoals}

At this time, STScI is looking to expand its DOI Pilot Project in the following ways:
\begin{itemize}
\item Expand E-Journal Press submission prompt to 18 additional institutions (mostly universities). Doing so will help MAST understand how a wider audience interacts with the data archive and improve its DOI service with feedback from the community. This also creates opportunities for discussion on the purpose and goals of data attribution in astronomy.
\item Implement a DOI minting service for data generated by research in addition to data analyzed during research. To do this, we need a defined standard for \textit{relatedIdentifier} values and specifically \textit{relationType} showing the association between data analyzed and data generated.\\

MAST is currently working with the AAS journals to expand the pilot to solicit data products from scientists at the time of paper submission. This will streamline the process for authors wishing to contribute data generated, ensure proper data linking within the manuscript, provide authors with a unique DOI to reliably refer to their MAST-hosted data product, and also benefit MAST by advertising this data hosting service to more authors.
\item Create a utility for "sample DOIs". As noted in \S \ref{sec:techchallenges}, many manuscripts use subsets of larger catalogs or atlases, such as uniform lists of galaxies or stars. At present the selection criteria are often described in words, and sometimes query-language is published within the manuscript. These methods are cumbersome and often incomplete. A "Sample DOI" which keeps track of these analyzed sub-samples during the research process, rather than lists of primary data observations, will be technically challenging, but valuable. This will require greater integration of HLSPs within the MAST Discovery Portal so researchers can sort and select sample data from these larger products and retain these subsets for later, permanent DOI creation.
\item Align the MAST metadata schema with DataCite Metadata Schema 4.1 guidelines, and more specifically ensure our existing DOIs have an added DataCite property for \textit{relatedIdentifierType} outlining the relationship between the data (DataCite DOI) and research article (CrossRef DOI), and/or negotiate with publishers so they define this relationship in the article (CrossRef) metadata upon publication.
\item Assist other archives with implementing a DOI service.
\item Allow for proper attribution via metadata to credit primary data creators, i.e., the principal investigators who proposed the initial observations, and also the archives that curate the data to ensure it is findable, accessible, interoperable, and reproducible (\emph{FAIR}). 
\item Consider minting DOIs for data sets already identified in earlier papers that are part of the HST bibliography and link to these DOIs via the data link feature in ADS. This is to avoid future broken links.
\item Explore partnerships with other major publishers (Oxford, EDP Sciences, etc.).
\item Develop a long-term preservation policy so the archive remains committed to providing resolvable DOIs, in the event that the archive were ever decommissioned or absorbed by another archive.
\item The ultimate goal of the DOI pilot project group at STScI is to fully support the use of DOI services within the MAST Portal for all authors using MAST data, regardless of affiliation, by the time the first scientific observations from JWST are delivered.

Creating a consistent data citation culture around JWST could greatly influence the field at large to do the same for other observatories.
\end{itemize}
    
\section{Conclusions} \label{sec:conclusions}

Higher standards in refereed journals for data citation, and integration of a DOI creation service within an astronomical archive will encourage authors to attribute data more frequently and with less ambiguity. Being able to associate publications with the data analyzed and data generated will allow future researchers to understand the methods and criteria applied to data analysis and derivation, and lay the foundation for a data citation and indexing culture in which those who are responsible for the creation of data are credited for their contributions.
\\  

\textit{``A publication based on a data set is just one expression of the potential in that data set. The backgrounds and interests of the researchers will influence which representation of that data is selected. But there are many different representations, and the ability to discover and access data products fosters the reuse of data products for different purposes as well as for combining data products in unanticipated combinations.''\\
---Edwin Henneken, ADS Technologist, 2015}

\acknowledgments
Our thanks go to Jill Lagerstrom, former Chief Librarian at Space Telescope Science Institute, who helped lay the foundation for data DOIs and their potential for integration with the HST Bibliography. The DOI pilot project group at STScI extends their thanks to Tom Donaldson, Amanda Marrione, and Randy Thompson at MAST. This group, along with many others, work tirelessly on MAST Portal integration and HLSP maintenance and make data discovery for the astronomy community possible. We also wish to thank the staff at ADS, including Alberto Accomazzi and Edwin Henneken for their ideas, insight, dedication, and support.

\end{document}